\documentclass[useAMS,usenatbib]{mn2e_x}

\usepackage{graphicx}
\usepackage{natbib}
\usepackage{graphicx}
\usepackage{color}
\usepackage{pdfpages}
\usepackage{appendix}
\usepackage{subfigure}
\usepackage{url}
\usepackage{amsmath}
\usepackage{float}

\begin{document}
\bibliographystyle{aabib}

\def\py{\textsc{Python}}
\def\tar{\textsc{Tardis}}
\def\cld{\textsc{Cloudy}}
\def\agn{\textsc{Agnspec}}

\def\civ{C~\textsc{iv}}
\def\nv{N~\textsc{v}}
\def\hei{He~\textsc{i}}
\def\heii{He~\textsc{ii}}
\def\heiiline{He~\textsc{ii}~$4686$\AA}
\def\mg{Mg~\textsc{ii}}
\def\al{Al~\textsc{iii}}
\def\heii{He~\textsc{ii}}
\def\ovi{O~\textsc{vi}}
\def\la{Ly~$\alpha$}
\def\ha{H$\alpha$}
\def\hb{H$\beta$}
\def\civline{C~\textsc{iv}~$1550$\AA}
\def\nvline{N~\textsc{v}~$1240$\AA}
\def\mgline{Mg~\textsc{ii}~$2800$\AA}

\def\araa{ARAA}
\def\nat{Nature}
\def\apjl{ApJ Letters}
\def\aapr{AAPR}
\def\ssr{SSR}
\def\apj{ApJ}
\def\apjs{ApJs}
\def\pasp{PASP}
\def\aap{A\&A}
\def\mnras{MNRAS}
\def\aj{AJ}
\def\rmxaa{RMXAA}
\def\aaps{A\&As}
\def\LA{Lyman\thinspace$\alpha$}

\newcommand{\EXPN}[2]{\mbox{$#1\times 10^{#2}$}}
\newcommand{\EXPU}[3]{\mbox{\rm $#1 \times 10^{#2} \rm\:#3$}}  
\newcommand{\POW}[2]{\mbox{$\rm10^{#1}\rm\:#2$}}
\def\LUM{\:{\rm erg\:s^{-1}}}
\def\FLUX{\:{\rm erg\:cm^{-2}\:s^{-1}}}
\def\OIGS{\:{\rm erg\:cm^{-2}\:s^{-1}\:\AA^{-1}}}

%
%

\title
[Radiative Transfer in Clumpy Winds]
{
Testing Quasar Unification: Radiative Transfer in Clumpy Winds
}

\author[Matthews et al.]{
\parbox[t]{\textwidth}{
J.~H.~Matthews$^1$\thanks{jm8g08@soton.ac.uk}, C.~Knigge$^1$, 
K.~S.~Long$^2$, S.~A.~Sim$^3$, N.~Higginbottom$^1$ and S.~W.~Mangham$^1$
}
\medskip  
\\$^1$School of Physics and Astronomy, University of Southampton, Highfield, Southampton, SO17 1BJ, United Kingdom
\\$^2$Space Telescope Science Institute, 3700 San Martin Drive, Baltimore, MD, 21218
\\$^3$School of Mathematics and Physics, Queens University Belfast, University Road, Belfast, BT7 1NN, Northern Ireland, UK
}

\date{Accepted 8 February 2016. Received 8 February 2016; in original form 23 December 2015}
%
%

\maketitle
\begin{abstract} 
Various unification schemes interpret the complex phenomenology of quasars and luminous active galactic nuclei (AGN) in terms of a simple picture involving a central black hole, an accretion disc and an associated outflow. Here, we continue our tests of this paradigm by comparing quasar spectra to synthetic spectra of biconical disc wind models, produced with our state-of-the-art Monte Carlo radiative transfer code. Previously, we have shown that we could produce synthetic spectra resembling those of observed broad absorption line (BAL) quasars, but only if the X-ray
luminosity was limited to 10$^{43}$~erg~s$^{-1}$. 
Here, we introduce a simple treatment of clumping, and find that a filling factor of $\sim0.01$ 
moderates the ionization state sufficiently for BAL features to form in the 
rest-frame UV at more realistic X-ray luminosities. Our fiducial model shows good agreement with AGN X-ray properties and the wind produces strong line emission in, e.g., \la\ and \civline\ at low inclinations. At high inclinations, the spectra possess prominent LoBAL features. Despite these successes, we cannot reproduce all emission lines seen in quasar spectra with the correct equivalent-width ratios, and we find an angular dependence of emission-line equivalent width despite the similarities in the observed emission line properties of BAL and non-BAL quasars. Overall, our work suggests that biconical winds can reproduce much of the qualitative behaviour expected from a unified model, but we cannot yet provide quantitative matches with quasar properties at all viewing angles. Whether disc winds can successfully unify quasars is therefore still an open question.
\end{abstract}


%
%

\section{Introduction}

The spectra of 
quasars and luminous active galactic nuclei (AGN) 
typically exhibit a series of strong emission lines
with an underlying blue continuum - the so-called {\sl `big blue bump'} (BBB). 
The BBB is often attributed to emission from a 
geometrically thin, optically thick accretion disc surrounding the central black hole (BH), similar to that described by \cite{shakurasunyaev1973}.
In addition to the {\em inflowing} accreting material, 
{\em outflows} are ubiquitous in AGN
and quasars \citep{kellerman1989,ganguly2008}. These outflows can take the form of 
highly collimated radio jets \citep[e.g.][]{hazard1963,potash1980,perley1984,marscher2006}, 
or mass-loaded `winds' emanating from the accretion disc 
\citep{weymann1991,turnermiller2009}. 
Outflows in AGN offer a 
potential feedback mechanism through which the central source can 
affect its environment \citep{king2003,king2005,fabian2012}
-- feedback that is required in models of galaxy evolution \citep{springel2005}
and may explain the `$M-\sigma$' relation \citep{silkrees1998,haring2004}.

Perhaps the clearest evidence of outflows in AGN is  
the blueshifted ($\sim 0.1c$) broad absorption lines (BALs) in the 
ultraviolet seen in approximately $20\%$ of quasars
\citep{weymann1991, knigge2008, allen2011}.
The simplest explanation for the incidence of 
BAL quasars (BALQSOs) is in terms of an accretion disc wind. 
According to this paradigm, a biconical wind rises from 
the accretion disc and the BALQSO fraction is associated with
the covering factor of the outflow. 
Polarisation studies suggest that the wind is roughly equatorial
\citep{goodrich1995, cohen1995}, although there is also evidence
for polar BAL outflows  in radio-loud (RL) sources \citep{zhou2006,ghoshpunsly2007}.

Due to their ubiquitous nature,
disc winds offer a natural way to {\em unify} much of the
diverse phenemonology of luminous AGN and quasars \citep[e.g.][]{MCGV95, elvis2000}. 
Depending on viewing angle, an observer 
may see a BALQSO or normal `Type 1' quasar.
Within this geometric unification framework, the broad-line region (BLR) can 
correspond either to the dense wind base or clumps embedded
in the outflow. 
A biconical wind model can also readily explain the various sub-classifications of BALQSOs: 
HiBALQSOs, which only exhibit high ionization line absorption; LoBALQSOs, which also show
absorption in lower ionization state species such as \mg\ and \al; and
FeLoBALQSOs, which show further absorption in Fe~\textsc{ii} and \textsc{iii}.
In unified geometric models, this is generally attributed to ionization stratification
of the outflow \citep[e.g.][]{elvis2000}.

Despite the importance of disc winds in shaping quasar and AGN spectra,  
much of the underlying outflow physics remains uncertain. 
Several driving mechanisms have been proposed, including
thermal pressure \citep{weymann1982, begelman1991}, magnetocentrifugal forces 
\citep{blandfordpayne,pelletier_pudritz} and 
radiation pressure on spectral lines \citep[`line-driving';][]{lucysolomon1970,shlosman1985,MCGV95}.
Of these, line-driving is possibly the most attractive, as
strong absorption lines are already seen in BALQSOs and the X-ray spectra of AGN 
\citep{reeves2003,poundsreeves2009,tombesi2010a}.
The efficiency of line-driving is crucially dependent on the ionization state 
of the outflow, and in AGN, unlike O stars, 
the presence of strong X-ray emission can overionize the wind, 
causing it to `fail'.
\cite{MCGV95} proposed a potential solution: 
a region of `hitchhiking gas' that could shield the wind from the central X-ray source. 
An additional or alternative solution is that the wind is clumped 
\citep[e.g.][]{junk1983,weymann1985,hamann2013}
possibly on multiple scale lengths. Local density enhancements could lower the 
ionization parameter of the plasma 
without requiring excessively large total column densities 
and mass-loss rates.

Evidence for dense substructures in AGN winds is widespread.
BALQSOs show complex absorption line profiles \citep{ganguly2006, simonhamann2010}
and exhibit variability in these profile shapes \citep{capellupo2011,capellupo2012,capellupo2014}.
AGN generally show variability in X-ray absorption components \citep[e.g.][]{risaliti2002}
and many models for the BLR consist of clumps embedded in an outflow 
\citep{krolik1981, emmering1992, dekool1995, cassidyraine1996}.
\cite{krolik1981} showed that BLR clouds would be short-lived, suggesting 
that a confining mechanism is required, such as
magnetic confinement \citep[e.g.][]{dekool1995}. Alternatively,
the `line deshadowing instability' (LDI) inherent in line-driven winds
\citep{lucysolomon1970,macgregor1979,carlberg1980,owockirybicki1984}
may be responsible for smaller-scale clumping, as is observed in O-star winds
\citep[][and references therein]{fullerton2011}.
Indeed, simple models of clumping have successfully explained the electron scattering 
wings of emission lines formed in these line-driven flows 
\citep{hillier1991eswingsmodel}.
Complex substructures are also produced in simulations of line-driven 
outflows in AGN, although on very different scales to LDIs 
\citep{PSK2000,PK04,progakurosawa2010,proga2014}.  
In summary, clumpy winds offer an observationally motivated and theoretically 
predicted way to lower the ionization state of a plasma, possibly in tandem
with a shielding scenario. However, it is equally important to note that there 
is no consensus view of the size or location of the clumps in the outflow.

We have been engaged in a project to determine whether it is possible to simulate the properties of 
the spectra of AGN, including BALQSOs, using simple kinematic prescriptions for biconical disc winds.
To address this question, we use a Monte Carlo radiative transfer (MCRT) code that calculates the ionization structure of the wind 
and simulates the spectra from such a system 
\citep[][hereafter H13 and H14]{simlong2008,sim2010,higginbottom2013,H14}.  The results have been encouraging in the sense that in H13, we showed we could produce simulated spectra that resembled that of BALQSOs, as long as the luminosity of the X-ray source was relatively low, of order \POW{43}{\LUM} and the mass loss rate was relatively high, of order the mass accretion rate.  However, at higher X-ray luminosities, the wind was so ionized that UV absorption lines were not produced.  In addition, and in part due to limitations in our radiative transfer code, the model failed to produce spectra with strong emission lines at any inclination angle.  

Here we attempt to address both of these issues, by introducing clumping into our model and a more complete treatment of H and He into our radiative transfer calculations.   
The remainder of this paper is organized as follows:
In section 2, we describe some of the important photoionization 
and MCRT aspects of the code. We then outline the model in section 3, including 
a description of our clumping implementation and success criteria. 
Section 4 contains the results from a clumped model, 
with comparisons to observational data, as well as some discussion. 
Finally, we summarise our findings in section 5.




\section{Ionization and Radiative Transfer}

For this study, we use the MCRT code \py 
\footnote{Named {\sl c. 1995}, predating the inexorable rise of a certain progamming language.} we have developed to carry out our 
radiative transfer and photoionization simulations in non-local-thermodynamic-equilibrium 
(non-LTE). The code can be used to model a variety of
disc-wind systems; applications have included accreting white dwarfs 
(Long \& Knigge 2002, hereafter LK02; Noebauer et al. 2010; 
Matthews et al. 2015, hereafter M15), young-stellar objects 
\citep{simmacro2005} and quasars/AGN (H13, H14).\nocite{noebauer, M15, LK02}  

The code operates as follows:   This outflow is discretized into $n_x \times n_z$ cells in a 2.5D
cylindrical geometry with azimuthal symmetry. 
From some initial conditions in each cell (typically $T_e \sim 40,000$K in AGN, with 
dilute LTE abundances), 
the code first calculates the ionization structure of the wind in a series of iterations. 
Each iteration in an ``ionization cycle'' consists of generating photons, actually photon packets, 
from an accretion disc and central object, and calculating how these photon 
bundles scatter through the wind (eventually escaping the outflow or hitting the disk). 
The ionization and temperature structure is updated based on the properties of the 
radiation field in each cell, and the process is repeated. The radiative transfer and thermal
balance is carried out according to the principle of radiative equilibrium, 
which assumes both statistical equilibrium and that radiative heating balances radiative cooling.
Once the ionization structure has converged, it is held fixed, 
and synthetic spectra are generated at specific inclination 
angles in a series of "spectral cycles". LK02 provide a more detailed  
description of the original code; various improvements have been made 
since then and are described by \cite{simmacro2005}, H13 and M15.  
We focus here on the specific changes made for this study
intended to improve the ionization calculation of H and He
and to allow for clumping in the wind.

\subsection{Line transfer}

Our approach to line transfer is based upon the macro-atom implementation developed by 
\cite{lucy2002, lucy2003}, in which the energy flows through the system are described in 
terms of indivisible energy quanta of radiant or kinetic energy 
(`$r-$packets' and `$k-$packets' respectively; see also section~\ref{sec:photon_sources}),
which are reprocessed by `macro-atoms' with associated transition probabilities.
We use the Sobolev approximation \citep[e.g.][]{sobolev1957,sobolev1960,rybickihummer1978}
to compute the location and optical depth of line interactions.
In our case, for reasons of computational efficiency, we adopt the  hybrid macro-atom scheme 
described by M15. In this scheme, the energy packets interact with either two-level 
`simple-atoms' or full macro-atoms. 
This allows one to treat non-LTE line transfer in radiative equilibrium
without approximation for elements that are identified as 
full macro-atoms, while maintaining the fast `two-level' 
treatment of resonance lines when elements are identified 
as simple atoms (see M15). In this study,
only H and He are treated as macro-atoms, because the process is computationally
intensive, and we expect recombination to be particularly important
in determining their level populations and resultant line emission.
We are also especially interested in the contribution to 
AGN spectra of \LA.  H13 treated all atoms in a two-level approximation.  

\subsection{Ionization and Excitation Treatment}

Macro-atoms have their ion and level populations derived from
MC rate estimators as described by Lucy (2002,2003). 
Previously (LK02, H13, M15), we used a modified Saha approach to 
calculate the ionization fractions
of simple-atoms. As part of  this effort, we have 
now improved {\sc PYTHON} to explicitly solve the 
rate equations between ions in non-LTE. This dispenses with a number of small assumptions 
made in the modified Saha approach, is more numerically stable, and
prepares for a more accurate treatment of other physical processes in future. 

In order to calculate the photoionization rate, 
we model the SED in a grid cell using the technique described by H13. In this scheme,
the mean intensity, $J_{\nu}$, in a series of $n$ bands is modeled as either a power law or exponential
in frequency $\nu$, with the fit parameters deduced from band-limited radiation field estimators.
This allows the calculation of a photoionization rate estimator. Ion abundances are
then calculated by solving the rate equations between ions. We include collisional ionization and photoionization balanced with radiative, 
dielectronic and collisional (three-body) recombination.
As in M15, we use a dilute Boltzmann approximation to calculate 
the populations of levels for simple-atoms. This dilute approximation 
is not required for macro-atom levels. 

\subsection{Physical Processes}

We include  free-free, bound-free and bound-bound heating
and cooling processes in the model. 
For radiative transfer purposes
we treat electron scattering in the Thomson limit, 
but take full account of Compton heating and cooling when
calculating the thermal balance of the plasma (see H13).
Adiabatic cooling in a cell is calculated using the estimator
\begin{equation}
C_{A} = f_V k_B T_e V (\nabla \cdot v) \left(n_e + \sum\limits_{i=1}^{ions} N_i\right),
\end{equation}
where $V$ is the cell volume, $T_e$ is the electron temperature,
$n_e$ is the electron density, $N_i$ is the number density of ion $i$
and $\nabla \cdot v$ is the divergence of the 
velocity field at the centre of the cell. The sum is over all ions 
included in the simulation.
For consistency, $k$-packets then have a destruction probability of 
\begin{equation}
P_{k,A} = \frac{C_{A}}{C_{tot}},
\end{equation}
where $C_{tot}$ is the sum over all cooling rates in the cell.

\subsection{Atomic Data}
We use the same atomic data  described by LK02 as updated by H13 and M15.
As in M15, macro-atom level information is from \textsc{Topbase} \citep{cunto1993}.
Photoionization cross-sections are from \textsc{Topbase} \citep{cunto1993} and \cite{vfky}.
Dielectronic and Radiative recombination rate coefficients for simple-atoms are taken from 
the \textsc{Chianti} database version 7.0 \citep{dere1997,landi2012}.
We use ground state recombination rates from \cite{badnell2006} where available,
and otherwise default to calculating recombination rates from the Milne
relation. Free-free Gaunt factors are from \cite{sutherland1998}.

%
%

\section{A Clumpy Biconical Disk Wind Model for Quasars}

\begin{figure} 
\centering
\includegraphics[width=0.45\textwidth]{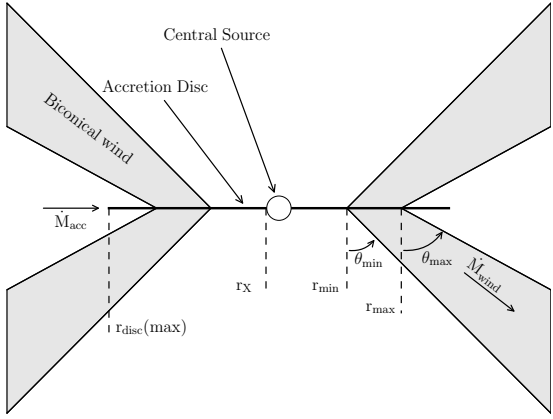}
\caption
{
A cartoon showing the geometry and some key parameters of
our biconical wind model.
}
\label{fig:cartoon}
\end{figure}

Our kinematic prescription for a biconical disc wind model
follows \cite{SV93}, and is described further by
LK02, H13 and M15. The purpose of this
purely kinematic wind model is to provide a simple tool for 
exploring different outflow geometries.
A schematic is shown in Fig.~\ref{fig:cartoon},
with key aspects marked. The general biconical
geometry is similar to that invoked by \cite{MCGV95} and 
\cite{elvis2000} to explain the phenomenonology
of quasars and BALQSOs. However, we do not include the initial
vertical component present in the Elvis (2000) model, which was
invoked as a qualitative explanation for narrow absorption lines.

\subsection{Photon Sources}
\label{sec:photon_sources}

We include two sources of r-packets in our model:
An accretion disc and a central X-ray source.
The accretion disc is assumed to be geometrically thin, but optically thick.
Accordingly, we treat the disc as an ensemble of blackbodies with a 
\cite{shakurasunyaev1973} effective temperature profile. 
The emergent SED is then determined by the specified accretion rate ($\dot{m}$)
and central BH mass ($M_{BH}$).
All photon sources in our model are opaque, meaning
that r-packets that strike them are destroyed.
The inner radius of the disc extends to the innermost 
stable circular orbit (ISCO) of the BH. 
We assume a Schwarzchild BH with an ISCO at $6~r_G$, where 
$r_G = GM_{BH}/c^2$ is the gravitational radius.
For a $10^9~M_\odot$ BH, this is equal to $8.8\times10^{14}~{\rm cm}$ 
or $\sim10^{-4}~{\rm pc}$.

The X-ray source is treated as an isotropic sphere at the ISCO,
which emits $r$-packets according to a power law in flux with index $\alpha_X$, of the form
\begin{equation}
F_X (\nu) = K_X \nu^{\alpha_X}.
\end{equation}
The normalisation, $K_X$ of this power law is such that it 
produces the specified 2-10~keV luminosity, $L_X$.
Photons, or $r$-packets, produced by the accretion disc and central X-ray source
are reprocessed by the wind. This reprocessing is dealt with by enforcing strict
radiative equilibrium ({\em modulo} adiabatic cooling; see section~2.3)
via an indivisible energy packet constraint (see Lucy 2002, M15).  

\subsection{Kinematics and Geometry}

In the SV93 model, a biconical disc wind rises from the accretion 
disc between launch radii $r_{min}$ and $r_{max}$.
The opening angles of the wind are set to $\theta_{min}$ and $\theta_{max}$.
The poloidal velocity along each individual streamline at a poloidal distance $l$ 
is then given by \citep[e.g.][]{rybickihummer1978}
\begin{equation}
v_l=v_0+\left[v_{\infty}(r_0)-v_0\right]\frac{\left(l/R_v\right)^{\alpha}}{\left(l/R_v\right)^{\alpha}+1},
\label{v_law}
\end{equation}
where $v_0$ is the velocity at the base of the streamline, $\alpha$ is
an exponent governing how quickly the wind accelerations and 
$R_v$ is the `acceleration length', defined as the distance at which
the outflow reaches half of its terminal velocity, $v_{\infty}$.
The terminal velocity is set to a fixed multiple of the escape
velocity, $v_{esc}$, at the base of the streamline (radius $r_0$).
The rotational velocity, $v_{\phi}$, is initially Keplerian ($v_k = [GM/r_0]^{1/2}$),
and the wind conserves specific angular momentum, such that 
\begin{equation}
v_{\phi} r = v_k r_0.
\label{v_law}
\end{equation}
The velocity law is crucial in determining the output spectra,
as it affects not only the projected velocities along the line of sight,
but also the density and ionization state of the outflow.
A wind that accelerates more slowly will have a denser wind base
with correspondingly different ionization and emission characteristics.

\subsection{A Simple Approximation for Clumping}

In our previous modelling efforts, we assumed a smooth outflow, 
in which the density at a given point was determined only by the 
kinematic parameters and mass loss rate. However, as already discussed,
AGN winds exhibit significant substructure -- the outflow is expected to be
{\em clumpy}, rather than smooth, and probably on a variety of scales. 
A clumpy outflow offers a possible solution to the so-called `over-ionization problem' in 
quasar and AGN outflows \citep[e.g.][]{junk1983,weymann1985,hamann2013}. 
This is the main motivation for incorporating clumping into our model.

Deciding on how to implement clumping into our existing wind models was not straightforward.
First, and most importantly, the physical scale lengths and density contrasts in AGN outflows are not well-constrained from observations or theory.  As a result, while one could envision in principle, clouds with a variety of sizes and density contrasts varying perhaps as function of radius, there would have been very little guidance on how to set nominal values of the various parameters of such a model.
Second, there are significant computational difficulties associated with adequately resolving and realistically modelling a series of small scale, high density regions with a MCRT
-- or for that matter, a hydrodynamical -- code. 
Given the lack of knowledge about the actual type of clumping, we have adopted
a simple approximation used successfully in stellar wind modelling, known as 
{\em microclumping} \citep[e.g.][]{hamann1998,hilliermiller1999,hamann2008}.  

The underlying assumption of microclumping is that clump sizes are much smaller than the 
typical photon mean free path, and thus the clumps are 
both geometrically and optically thin. This approach 
allows one to treat clumps only in terms of their volume filling factor, $f_V$, 
instead of having to specify separately their size and density distributions.
In our model, $f_V$ is independent of position (see section~4.4).
The inter-clump medium is modeled as a vacuum,
although the outflow is still non-porous and axisymmetric.
This approach therefore assumes that the inter-clump medium
is unimportant in determing the output spectrum, which
we expect to be true only when density constrasts are large and
the inter-clump medium is both very ionized and of low emissivity and opacity.
The density of the clumps is multiplied by the ``density enhancement'' 
$D=1/f_V$. Opacities, $\kappa$, and emissivities, $\epsilon$, 
can then be expressed as 
\begin{equation}
\kappa = f_V \kappa_C(D);~~\epsilon = f_V \epsilon_C(D).
\end{equation}
Here the subscript $C$ denotes that the quantity is calculated using the 
enhanced density in the clump. The resultant effect is that, {\em for fixed temperature},
processes that are linear in density, such as electron scattering, are unchanged, 
as $f_V$ and $D$ will cancel out. However, any quantity that scales with the square of density, 
such as collisional excitation or recombination, will increase by a factor of $D$.
In our models, the temperature is not fixed, and is instead set by balancing heating and 
cooling in a given cell. In the presence of an X-ray source, this thermal balance is 
generally dominated by bound-free heating and line cooling. The main effect of including 
clumping in our modelling is that it moderates the ionization state due to the increased 
density. This allows an increase in the ionizing luminosity, amplifying the amount of
bound-free heating and also increasing the competing line cooling term
(thermal line emission).

The shortest length scale in a Sobolev MCRT treatment such as ours 
is normally the Sobolev length, given by
\begin{equation}
l_S = \frac{v_{th}}{| dv/ds |}
\end{equation}
This is typically $\sim10^{13}$~cm near the disc plane, increasing outwards.
We use the mean density to calculate the Sobolev optical depth, which assumes that
$l_S$ is greater than the typical clump size.
Thus for the microclumping assumption to be formally correct, 
clumps should be no larger than $\sim10^{12}$~cm.
This size scale is not unreasonable for quasar outflows, as
\cite{dekool1995} suggest that BAL flows may have low filling factors with
clump sizes of $\sim10^{11}$~cm.

Our clumping treatment is necessarily simple; it does not adequately
represent the complex substructures and stratifications in ionization
state we expect in AGN outflows. 
Nevertheless, this parameterization 
allows simple estimates of the effect clumping has on the ionization 
state and emergent line emission.

\subsection{The Simulation Grid}

Using this prescription, we conducted a limited parameter
search over a 5-dimensional parameter space involving the 
variables $r_{min}$, $\theta_{min}$, $f_V$, $\alpha$ and $R_v$.
The grid points are shown in Table 1.
The aim here was to first fix $M_{BH}$ and $\dot{m}$ to their H13 values,
and increase $L_X$ to $10^{45}$~erg~s$^{-1}$ (a more realistic value for a 
quasar of $10^9M_\odot$ and an Eddington fraction of $0.2$; see section~\ref{sec:xray}).

We then evaluated these models based on 
how closely their synthetic spectra reproduced the 
following properties of quasars and BALQSOs:

\begin{itemize}
\item UV absorption lines 
with $BI > 0$ at $\sim20\%$ of viewing angles (e.g. Knigge et al. 2010);
\item Line emission emerging at low inclinations, with $EW\sim40$\AA\ in \civline\ \citep[e.g. ][]{shen2011};
\item H recombination lines with $EW\sim50$\AA\ in \la\ \citep[e.g. ][]{shen2011};
\item  \mg\ and \al\ (LoBAL) absorption features with $BI > 0$ at a subset of 
BAL viewing angles;
\item Verisimilitude with quasar composite spectra.
\end{itemize}
Here $BI$ is the `Balnicity Index' (Weymann et al. 1991), given by
\begin{equation}
BI = \int^{25000~{\rm km~s}^{-1}}_{3000~{\rm km~s}^{-1}} \left( 1 - \frac{f(v)}{0.9} \right) dv.
\end{equation}
The constant $C=0$ everywhere, unless the normalized flux
has satisfied $f(v)<0.9$ continuously for at least $2000$~km~s$^{−1}$, 
whereby $C$ is set to $1$.

In the next section, we present one of the most promising models,
which we refer to as the fiducial model, and discuss
the various successes and failures with respect to the above criteria.
This allows us to gain insight into fundamental geometrical 
and physical constraints and assess the potential for unification. 
We then discuss the sensitivity to key parameters in section~\ref{sec:param_sens}.
The full grid, including output synthetic spectra and plots can be found at
\url{jhmatthews.github.io/quasar-wind-grid/}.

\begin{table}
\begin{tabular}{p{2cm}p{1cm}p{1cm}p{1cm}p{1cm}}
Parameter & \multicolumn{4}{|l|}{Grid Point Values}  \\
\hline \hline 
$r_{min}$ 	&	 $60r_{g}$ & $180r_{g}$ & \multicolumn{2}{|l|}{$300r_{g}$} \\ 
$\theta_{min}$ 	& $55^{\circ}$ & \multicolumn{3}{|l|}{$70^{\circ}$} \\ 
$R_v$  	        &	 $10^{18}$cm & \multicolumn{3}{|l|}{$10^{19}$cm} \\ 
$\alpha$ 	&	 $0.5$ & $0.6$ & $0.75$ & $1.5$ \\
$f_V$ 	&	 $0.01$ & \multicolumn{3}{|l|}{$0.1$}  \\
\hline 
\end{tabular}
\caption{The grid points used in the parameter search.
The sensitivity to some of these parameters is discussed 
further in section~\ref{sec:param_sens}}
\label{grid_table}
\end{table}




\section{Results and Discussion From A Fiducial Model}
Here we describe the results from a fiducial model,
and discuss these results in the context of the criteria 
presented in section 3.4. The parameters of this model are shown in Table~2.
Parameters differing from the benchmark model of H13 are 
highlighted with an asterisk. In this section, we examine the physical 
conditions of the flow, and present the synthetic spectra, before comparing
the X-ray properties of this particular model to samples of
quasars and luminous AGN. 

\begin{table}
\begin{tabular}{p{3cm}p{4cm}}
\hline Fiducial Model Parameters 	&	 Value \\ 
\hline \hline 
$M_{BH}$ 	 &	 $1\times 10^9~\rm{M_{\odot}}$ \\ 
$\dot{m}_{acc}$ 	 &	 $5~M_{\odot}yr^{-1} \simeq 0.2~\dot{M}_{Edd}$\\ 
$\alpha_X$ 	 &	 $-0.9$ \\ 
$L_{X} $ 	 &	 $10^{45}~\rm{erg~s^{-1}}$$^*$ \\ 
$r_{disc}(min)=r_{X}$   &	 $6r_g=8.8\times10^{14}~{\rm cm}$ \\ 
$r_{disc}(max)$   &	 $3400r_g = 5\times10^{17}~{\rm cm}$ \\ 
$\dot{m}_{wind}$  &	 $5~M_{\odot}yr^{-1}$ \\ 
$r_{min}$ 	&	 $300r_{g} = 4.4\times10^{16}~{\rm cm}$\\ 
$r_{max}$ 	&	 $600r_{g} = 8.8\times10^{16}~{\rm cm}$ \\ 
$\theta_{min}$ 	&	 $70.0^{\circ}$ \\ 
$\theta_{max}$ 	&	 $82.0^{\circ}$ \\ 
$v_{\infty}(r_0)$ 	&	 $v_{esc}(r_0)$ \\ 
$R_v$  	        &	 $10^{19}$cm$^*$ \\ 
$\alpha$ 	&	 $0.5^*$ \\
$f_V$ 	&	 $0.01^*$  \\
$n_x$ 	&	 $100$  \\
$n_z$ 	&	 $200$  \\
\end{tabular}
\caption{Wind geometry parameters 
used in the fiducial model, as defined in the text and figure 1.
Parameters differing from the benchmark model of H13 are 
highlighted with an asterisk.}
\label{wind_param}
\end{table}

\subsection{Physical Conditions and Ionization State}

\begin{figure*}
\centering
\includegraphics[width=1.0\textwidth]{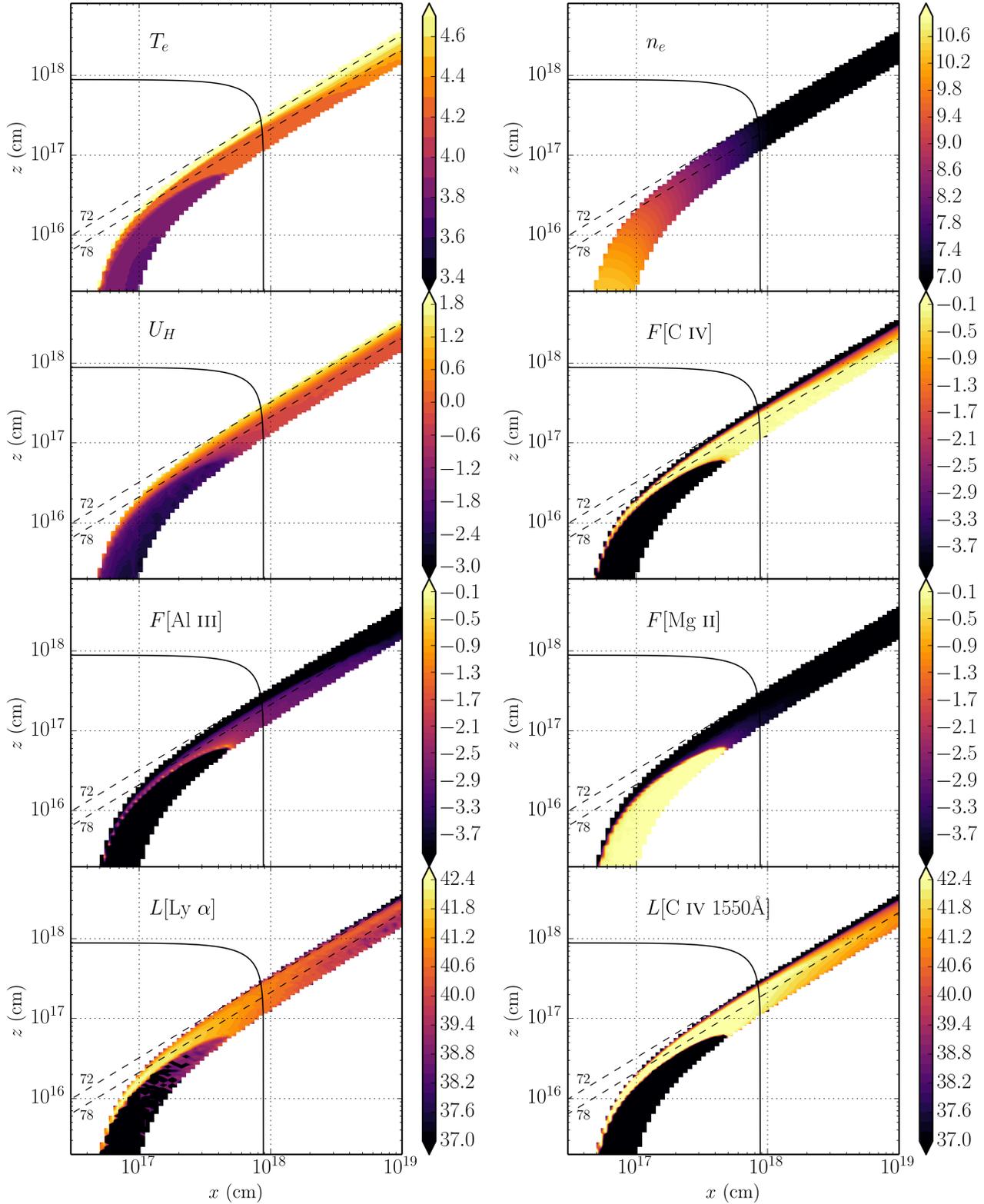}
\caption
{
Contour plots showing the logarithm of some important 
physical properties of the outflow. The spatial scales are
logarithmic and the $x$ and $z$ scales are not the same.
Symbols are defined in the text.
The solid black line marks a sphere at $1000~r_G$.
The dotted lines show the $72^\circ$ and $78^\circ$ sightlines 
to the centre of the system, and illustrate that different sightlines
intersect material of different ionization states.
The line luminosities, $L$, represent the luminosity of photons
escaping the Sobolev region for each line. These photons do not
necessarily escape to infinity.
}
\label{fig:wind}
\end{figure*}

\noindent
Fig.~\ref{fig:wind} shows the physical properties of the wind.
The wind rises slowly from the disc at first, with densities within clumps
of $n_H \sim 10^{11}~\rm{cm^{-3}}$ close to the disc plane, 
where $n_H$ is the local number density of H.
The flow then accelerates over a scale length of $R_V=10^{19}~\rm{cm}$
up to a terminal velocity equal to the escape velocity at the streamline base
($\sim10,000~\rm{km~s^{-1}}$). This gradual acceleration results in
a wind that exhibits a stratified ionization structure, with low ionization material
in the base of the wind giving way to highly ionized plasma further out.
This is illustrated in Fig.~\ref{fig:wind} 
by the panels showing the ion fraction $F=n_j/n_{tot}$ of some important ions.
With a clumped wind, we are able to produce the range of ionization states observed
in quasars and BALQSOs, while adopting a realistic $2-10$ keV X-ray luminosity
of $L_{X}=10^{45}~\rm{erg~s^{-1}}$. Without clumping, this wind would be over-ionized 
to the extent that opacities in e.g., \civ\ would be entirely negligible (see H13).

One common way to quantify the ionization state of a plasma
is through the ionization parameter, $U_H$, given by
\begin{equation}
U_H = \frac{4\pi}{n_H c}\int_{13.6{\rm{eV}} / h}^{\infty}\frac{{J_{\nu}d\nu}}{h\nu}.
\end{equation}
\noindent where $\nu$ denotes photon 
frequency. Shown in Fig.~\ref{fig:wind},
the ionization parameter is a useful measure of the global ionization state,
as it represents the ratio of the number density of 
H ionizing photons to the local H density.
It is, however, a poor representation of the 
ionization state of species such as \civ\ as it encodes no information
about the shape of the SED. In our case, the X-ray photons 
are dominant in the photoionization of the UV resonance line ions. 
This explains why a factor of 100 increase in X-ray luminosity requires
a clumping factor of 0.01, even though the value of $U_H$ decreases by only a factor of $\sim10$ compared to H13. 

The total line luminosity also increases dramatically compared to the unclumped model
described by H13. This is because the denser outflow can absorb the increased
X-ray luminosity without becoming over-ionized, leading to a hot plasma which
produces strong collisionally excited line emission.
This line emission typically emerges on the edge of the wind
nearest the central source. The location of the line emitting regions
is dependent on the ionization state, as well as the incident X-rays.
The radii of these emitting regions is important,
and can be compared to observations. The line luminosities, $L$,
shown in the figure correspond to the luminosity in $\LUM$ of photons
escaping the Sobolev region for each line. 
As shown in Fig.~\ref{fig:wind},
the \civline\ line in the fiducial model is typically formed between 
$100-1000~r_G$ ($\sim10^{17}-10^{18}~\rm{cm}$).
This is in rough agreement with the reverberation mapping 
results of Kaspi (2000) for the $2.6\times10^{9} M_\odot$ quasar S5 0836+71,
and also compares favourably with microlensing measurements of the size of the
\civline\ emission line region in the BALQSO H1413+117 \citep{odowd2015}.

\subsection{Synthetic Spectra: Comparison to Observations}

\begin{figure*}
\centering
\includegraphics[width=1.0\textwidth]{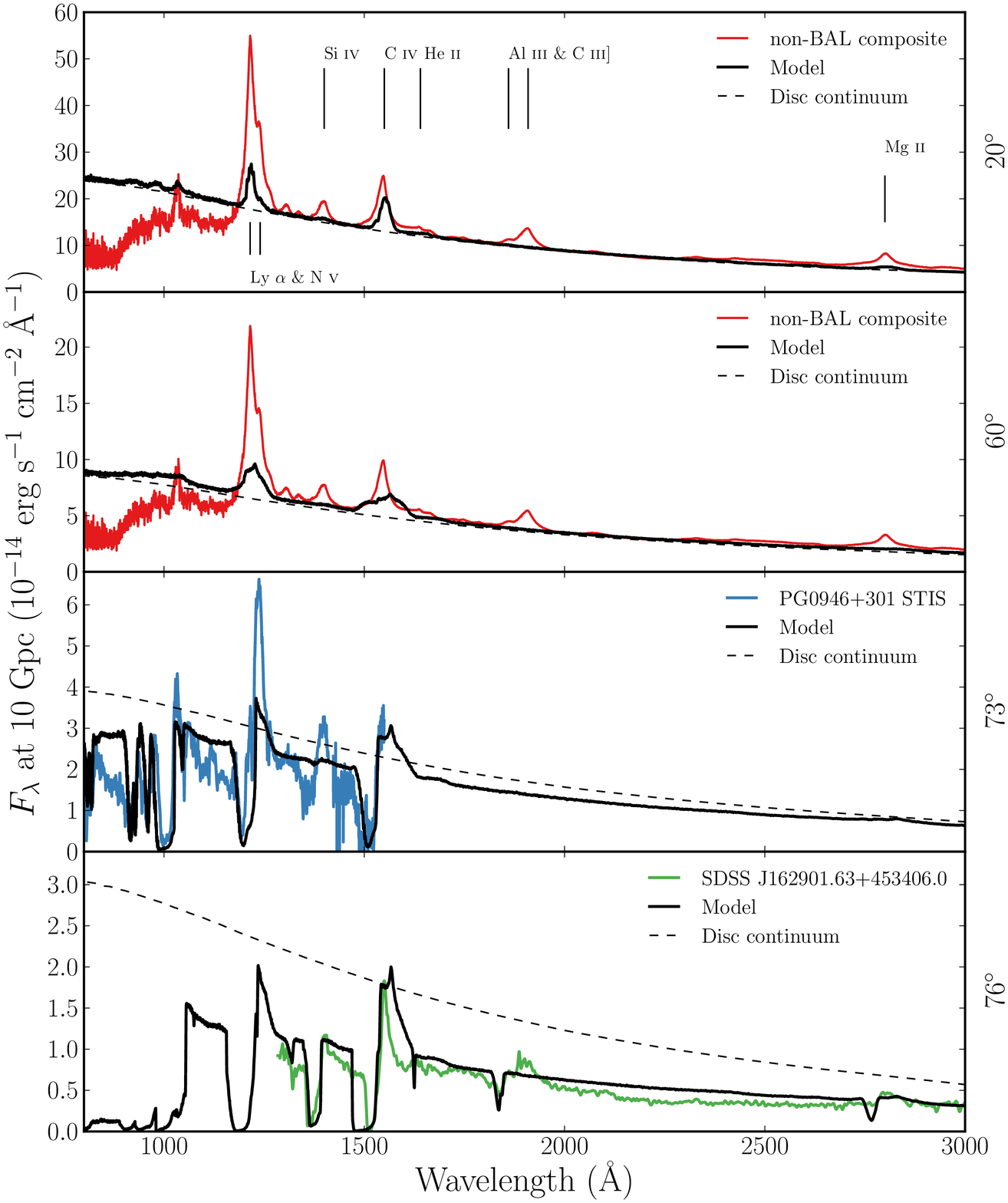}
\caption
{
Synthetic spectra at four viewing angles for the fiducial model. At 
$20^\circ$ and $60^\circ$ we show a comparison to an SDSS quasar composite
from Recihard et al. (2003). At $73^\circ$ and $76^\circ$ we show a comparison to
an {\sl HST} STIS spectrum of the high BALnicity BALQSO 
PG0946+301 (Arav et al. 2000), and an SDSS spectrum of the LoBAL quasar 
SDSS J162901.63+453406.0, respectively. The dotted line shows a disc
only continuum to show the effect of the outflow on the continuum level. 
All the spectra are scaled to the model flux at $2000$\AA, expect for the 
{\sl HST} STIS spectrum of PG0946+301, which is scaled to $1350$\AA\
due to the incomplete wavelength coverage.
}
\label{fig:uvspec}
\end{figure*}

\begin{figure} 
\centering
\includegraphics[width=0.5\textwidth]{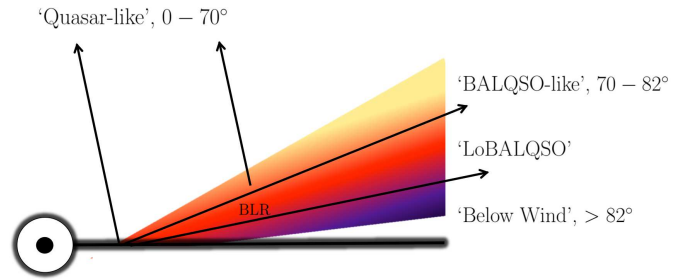}
\caption
{
A cartoon describing the broad classes of sightline 
in the fiducial model, illustrating how geometric effects lead to 
the different emergent spectra. The colour gradient is approximate,
but indicates the stratified ionization structure, 
from highly ionized (yellow) to low ionization (purple) material.
}
\label{fig:sightline}
\end{figure} 

\noindent
Fig.~\ref{fig:uvspec} shows the synthetic spectrum in the UV from the fiducial model. 
To assess the ability of the synthetic spectra to match real 
quasar spectra, we also show {\sl Sloan Digital Sky Survey} (SDSS) quasar
composites from \cite{reichard2003}, normalised to the flux at 2000\AA\
for low inclinations. Unfortunately, the wide variety of
line profile shapes and internal trough structure in BALQSOs
tends to `wash out' BAL troughs in composite spectra
to the extent that BALQSO composites do not resemble typical BALQSOs.
Because of this, we instead compare to a {\sl Hubble Space Telescope} 
STIS spectrum of the high BALnicity BALQSO PG0946+301 (Arav et al. 2000),
and an SDSS spectrum of the LoBAL quasar SDSS J162901.63+453406.0,
for the angles of $73^\circ$ and $76^\circ$, respectively. 
We show a cartoon illustrating how geometric effects determine
the output spectra in Fig.~\ref{fig:sightline}.  

\subsubsection{Broad absorption lines (`BALQSO-like' angles)}

The UV spectrum is characterised by strong BAL 
profiles at high inclinations ($> 70^\circ$). 
This highlights the first success of our model: 
clumping allows the correct ionization state 
to be maintained in the presence of strong X-rays, 
resulting in large resonance line opacities. 
At the highest inclinations, the 
cooler, low ionization material at the base of the wind
starts to intersect the line of sight. This produces 
multiple absorption lines in species such as \mg,
\al\ and Fe~\textsc{ii}. The potential links to LoBALQSOs and 
FeLoBALQSOs are discussed in section 2.4.

The high ionization BAL profiles are often saturated, and the location in velocity space
of the strongest absorption in the profile varies with inclination.
At the lowest inclination BAL sight lines, the strongest absorption occurs at the red edge,
whereas at higher inclinations (and for the strongest BALs)
the trough has a sharp edge at the terminal velocity.
This offers one potential explanation for the wide range of BALQSO absorption
line shapes (see e.g. Trump et al. 2006; Knigge et al 2008, Filiz Ak et al. 2014).

The absorption profiles seen in BALQSOs are often non-black, but saturated, 
with flat bases to the absorption troughs \citep{arav1999b,arav1999a}.
This is usually explained either as  partial covering of the continuum
source or by scattered contributions to the BAL troughs, necessarily
from an opacity source not co-spatial with the BAL forming region.
The scattered light explanation is supported by spectropolarimetry results
\citep{lamy2000}. Our spectra do not show non-black, saturated profiles.
We find black, saturated troughs at angles $i > 73^\circ$, and the BALs
are non-saturated at lower inclinations. The reasons for this are inherent 
in the construction of our model. 
First, the microclumping assumption does not allow for 
porosity in the wind, meaning that it does not naturally produce
a partial covering absorber. To allow this, an alternative approach
such as {\em macroclumping} would be required \citep[e.g.][]{hamann2008,surlan2012}.
Second, our wind does not have a significant scattering contribution 
along sightlines which do not pass through the BAL region,
meaning that any scattered component to the BAL troughs is absorbed by line opacity.
This suggests that either the scattering cross-section of the wind must
be increased (with higher mass loss rates or covering factors), or 
that an additional source of electron opacity is required, potentially
in a polar direction above the disc. We note the scattering contribution
from plasma in polar regions is significant in some `outflow-from-inflow'
simulations \citep{KP09, simproga2012}.

\subsubsection{Broad emission lines (`quasar-like' angles)}

\begin{figure*}
\centering
\includegraphics[width=1.0\textwidth]{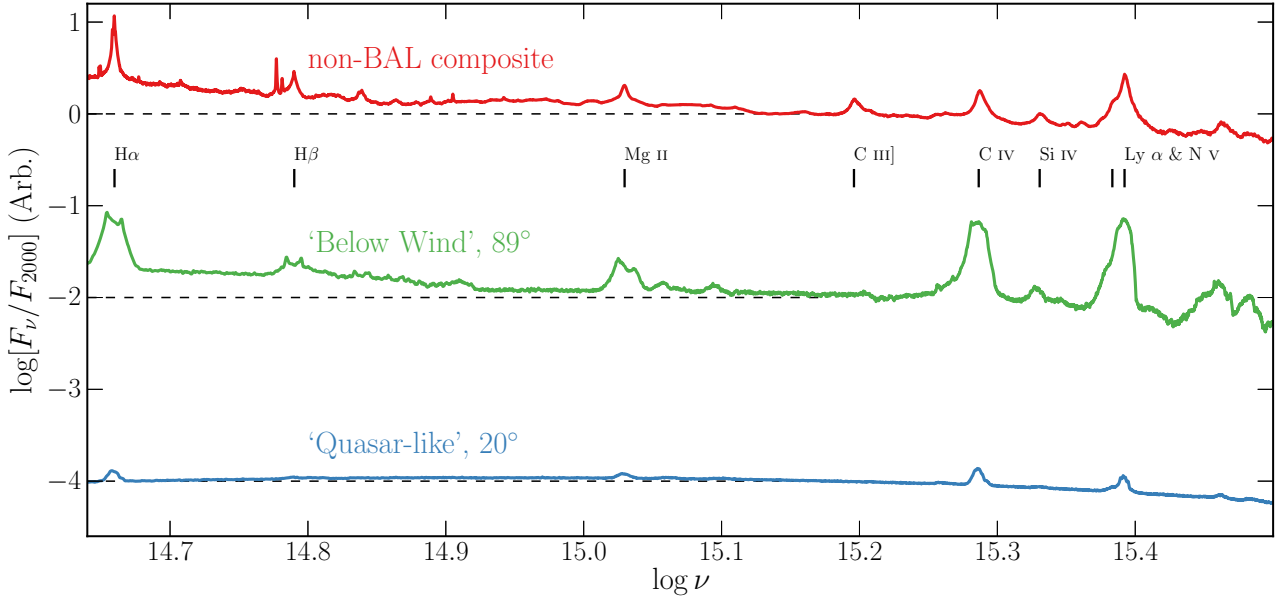}
\caption
{
Synthetic spectra at two viewing angles, 
this time in frequency space and including the optical band,
compared to the non-BAL SDSS quasar composite. The spectra are normalised to the flux at 
$2000$\AA, then an offset of 2 is applied per spectrum for clarity -- the dotted lines show the zero point of $F_\nu / F_{2000}$ in each case.
}
\label{fig:sed}
\end{figure*}

\begin{table}
\begin{tabular}{p{2cm}p{2cm}p{3cm}}
\hline Property & Synthetic, $20^\circ$ & Observed  (S11) \\ 
\hline \hline
$\log L$[C~\textsc{iv}]  & $44.60$ & $44.42 \pm 0.32$  \\
$\log L$[Mg~\textsc{ii}] & $43.92$ & $43.54 \pm 0.28$  \\
$\log (\nu L_{\nu})_{1350}$  & $46.42$ & $46.01 \pm 0.30$ \\
$\log (\nu L_{\nu})_{3000}$  & $46.18$ & $45.79 \pm 0.30$ \\
\hline
\end{tabular}
\caption{
Some derived spectral properties of the fiducial model, at $20^\circ$,
compared to observations. The observed values are taken from the Shen et al. (2011)
SDSS DR7 Quasar catalog, and correspond to mean values with standard deviations in log space
from a subsample with $8.5>\log(M_{BH})<9.5$ and $1.5<\log (L_{bol}/L_{Edd}) < 0$,
where the BH mass is a \civ\ virial estimate. 
Units are logarithms of values in erg~s$^{-1}$.
}
\label{line_lums}
\end{table}

Unlike H13, we now find significant collisionally excited line emission emerges
at low inclinations in the synthetic spectra, particular in the \civ\ and \nv\
lines. We also find a strong \la\ line and weak He~\textsc{ii}~$1640$\AA\ line
as a result of our improved treatment of recombination using macro-atoms. 
In the context of unification, this is a promising result, 
and shows that a biconical wind can produce significant 
emission at `quasar-like' angles. To demonstrate this further,
we show line luminosities and monochromatic continuum luminosities
from the synthetic spectra in Table~\ref{line_lums}. These are compared to
mean values from a subsample of the SDSS DR7 quasar catalog (Shen et al. 2011) 
with BH mass and Eddington fraction estimates similar to the fiducial model values 
(see caption). The spectra do not contain the strong 
C~\textsc{iii}]~1909\AA\ line seen in the quasar composite spectra, 
but this is due to a limitation of our current treatment of C; semi-forbidden
(intercombination) lines are not included in our modelling.

In Fig.~\ref{fig:sed}, we show an $F_{\nu}$ spectrum with broader waveband coverage
that includes the optical, showing that our synthetic spectra 
also exhibit \ha\ and \hb\ emission. 
In this panel, we include a low inclination and 
also a very high inclination 
spectrum, which looks underneath the wind cone. This model shows 
strong line emission with very similar widths and line ratios to the quasar composites, and
the Balmer lines are double peaked, due to velocity projection effects.  
Such double-peaked lines are seen in so-called `disc emitter' systems 
\citep[e.g.][]{eracleous1994} but not the majority of AGN.     
The line equivalent widths (EWs) increase at high inclination
due to a weakened continuum from wind attenuation, 
disc foreshortening and limb darkening. This effect also 
leads to a redder continuum slope, as seen in quasars, which is
due to Balmer continuum and Balmer and Fe~\textsc{ii} line emission.

The $89^\circ$ viewing angle cannot represent a typical quasar within a unified model, as
extreme inclinations should be extremely underrepresented in quasar samples.
This is in part due to the compton-thick absorber or `torus' 
expected at high inclinations \cite[e.g.][]{antonucci1985,martinez2007}.
However, this spectrum does show that a wind model can
naturally reproduce quasar emission lines if the emissivity of the wind is 
increased {\em with respect to the disc continuum}.
In addition, it neatly demonstrates how a stratified outflow can naturally
reproduce the range of ionization states seen in quasars.

Despite a number of successes, 
there are a some properties of the synthetic spectra
that are at odds with the observations. First, the ratios of the 
EW of the \la\ and \mgline\ lines
to the EW of \civline\ are much lower than in the composite spectra. 
Similar problems have also been seen in simpler photoionization models for the 
BLR \citep{netzer1990}.
It may be that a larger region of very dense ($n_e\sim10^{10}$cm$^{-3}$) 
material is needed, which could correspond to a disc atmosphere or 
`transition region' 
(see e.g. Murray et al. 1995; Knigge et al. 1997). \nocite{knigge1998} 
While modest changes to geometry may permit this, the initial grid search 
did not find a parameter space in which the \la\ or \mg\ EWs
were significantly higher (see section~\ref{sec:param_sens}). 
Second, we find that EWs increase with inclination 
(see Fig.~\ref{fig:uvspec} and Fig.~\ref{fig:sed}; also Fig.~\ref{fig:lobal}), 
to the extent that, even though significantly denser
models can match the line EWs fairly well at low inclinations, they will then
possess overly strong red wings to the BAL P-Cygni profiles at high inclinations.
The fact that the EW increase in our model are directly related to limb-darkening 
and foreshortening of the continuum. 
This appears to contradict observations, which show remarkably uniform emission
line properties in quasars and BALQSOs \citep{weymann1991,dipompeo2012b}. 
The angular distribution of the disc 
continuum and line emission is clearly crucially important in determining the emergent broad 
line EWs, as suggested by, e.g., the analysis of \cite{risaliti2011}. 
We shall explore this question further in a future study.  

\subsection{X-ray Properties}
\label{sec:xray}

\begin{figure*}
\centering
\includegraphics[width=0.9\textwidth]{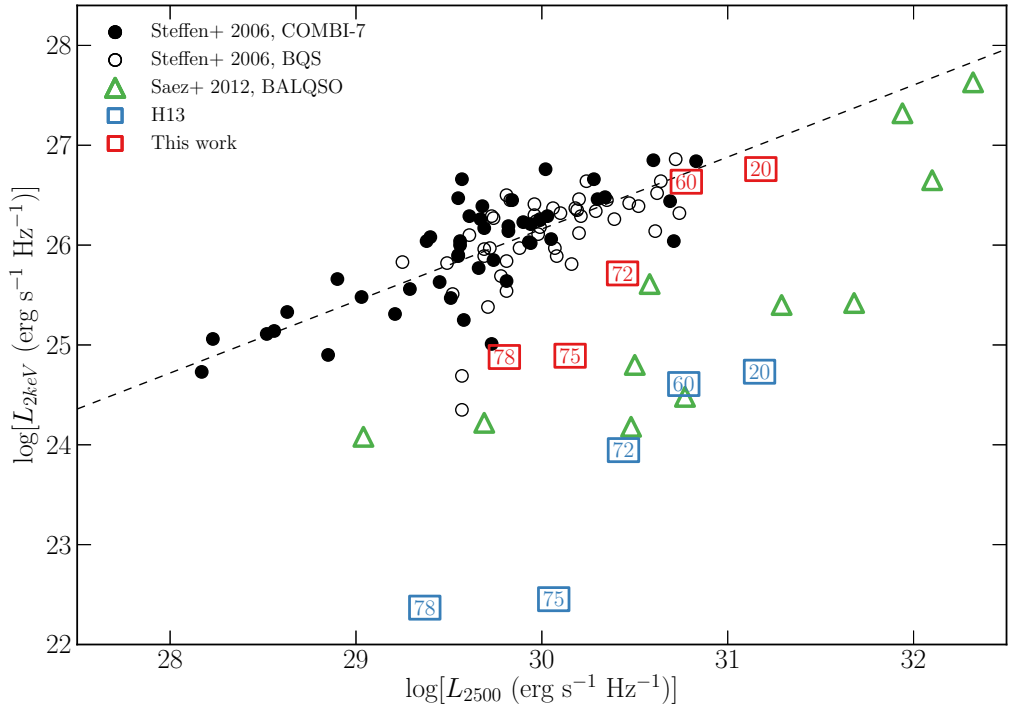}
\caption
{
X-ray ($2$~keV) luminosity of the our clumped model (red squares) 
and the H13 model (blue squares), plotted against monochromatic luminosity 
at 2500\AA. The points are labeled according to inclination; angles
$>70^\circ$ correspond to BALs in our scheme (see figure 4).
Also plotted are masurements from 
the COMBI-7 AGN and the BQS samples (Steffen et al. 2006) and the Saez et al. (2012) 
sample of BALQSOs. The dotted line shows the best fit relation for non-BALQSOs 
from Steffen et al. (2006).
}
\label{fig:xray}
\end{figure*}

The main motivation for adding clumping to the model was
to avoid over-ionization of the wind in the presence of strong X-rays. 
Having verified that strong BALs appear in the synthetic spectra,
it is also important to assess whether the X-ray properties of this
fiducial model agree well with quasar and BALQSO samples for the relevant
inclinations.

Fig.~\ref{fig:xray} shows the emergent
monochromatic luminosity ($L_\nu$) at 2~keV and 
plotted against $L_\nu$ at $2500$\AA\ for a number of different viewing angles in our model.
The monochromatic luminosities are calculated from the synthetic spectra and thus include
the effects of wind reprocessing and attenuation. In addition to model outputs,
we also show the BALQSO sample of Saez et al. (2012) and luminous AGN and quasar
samples from Steffen et al. (2006). The best fit relation from Steffen et al. (2006) 
is also shown. For low inclination, `quasar-like' viewing angles,
we now find excellent agreement with AGN samples. The slight gradient from $20^\circ$ to
$60^\circ$ in our models is caused by a combination of disc foreshortening and limb-darkening 
(resulting in a lower $L_{2500}$ for higher inclinations), and the fact that the disk 
is opaque, and thus the X-ray source subtends a smaller solid angle at high inclinations
(resulting in a lower $L_{2keV}$ for higher inclinations).

The high inclination, `BALQSO-like' viewing angles show moderate agreement with the data,
and are X-ray weak due to bound-free absorption and electron scattering in the wind.
Typically, BALQSOs show strong X-ray absorption with columns 
of $N_H\sim10^{23}~\rm{cm^{-2}}$ 
\citep{green1996,mathur2000,green2001,grupemathur2003}.
This is often cited as evidence that the BAL outflow is shielded from
the X-ray source, especially as sources with strong X-ray absorption tend
to exhibit deep BAL troughs and high outflow velocities 
\citep{brandt2000,laorbrandt2002,gallagher2006}.
Our results imply that the clumpy BAL outflow
itself can be responsible for the strong X-ray absorption, 
and supports Hamann et al.'s (2013) suggestion that 
geometric effects explain the weaker X-ray absorption in mini-BALs 
compared to BALQSOs.

\subsection{LoBALs and Ionization Stratification}

\begin{figure*}
\centering
\includegraphics[width=1.0\textwidth]{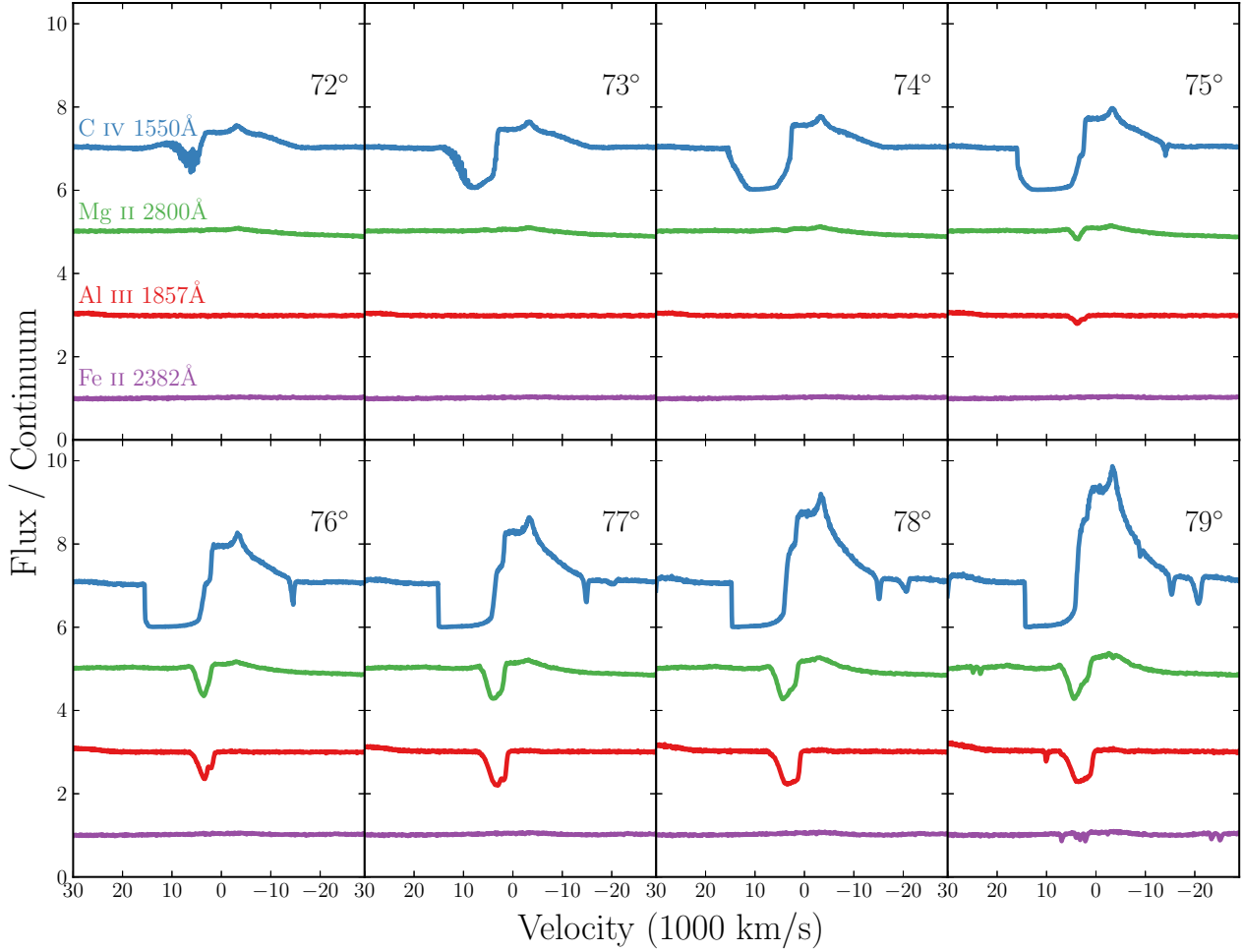}
\caption
{
\civ , \mg , \al\ and Fe~\textsc{ii} line profiles for viewing angles
from $72-79^\circ$. The profiles are plotted relative to the local
continuum with an offset applied for clarity. Lower ionization
profiles appear at a subset of high inclinations, compared
to the ubiquitous \civ\ profile.
}
\label{fig:lobal}
\end{figure*}

\begin{figure*}
\centering
\includegraphics[width=1.0\textwidth]{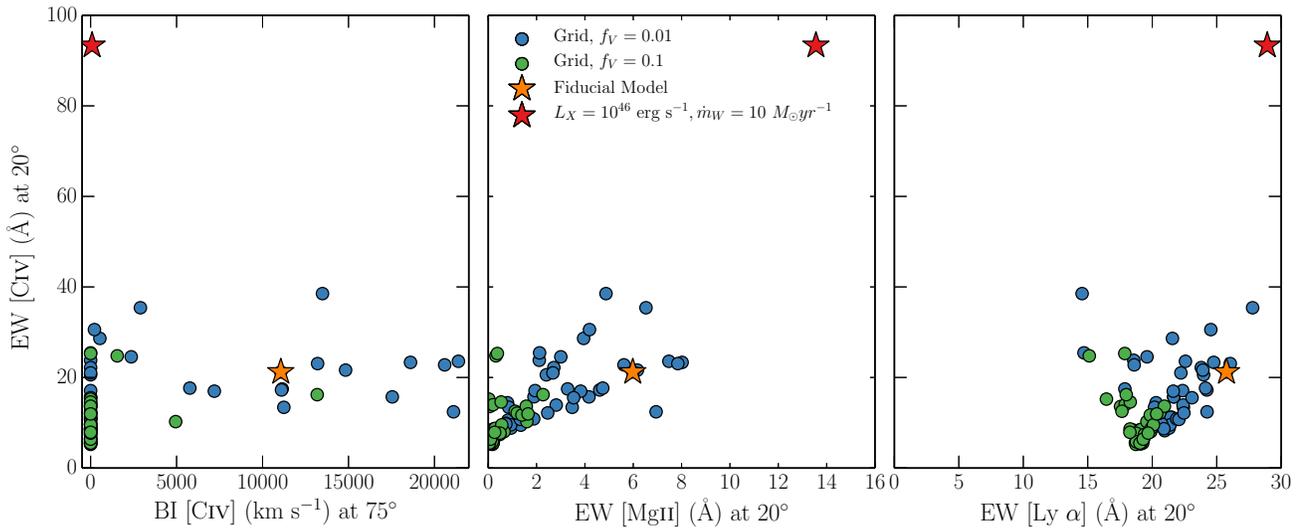}
\caption
{
The EW of the \civ~$1550$\AA\ line at $20^\circ$ plotted against a) the 
$BI$ of \civ~$1550$\AA\ at $75^\circ$, b) the EW of the \mg~$2800$\AA\ line 
at $20^\circ$ and c) the EW of \la\ at $20^\circ$. The circles correspond 
to the simulation grid for two different values of $f_V$, and the fiducial 
model is marked with an orange star. 
We also show a higher X-ray luminosity model and a higher mass loss rate
with a red star.
}
\label{fig:grid}
\end{figure*}

At high inclinations, the synthetic spectra exhibit blue-shifted BALs in \al\ and \mg --
the absorption lines seen in LoBALQSOs, and we even see absorption in Fe~\textsc{ii}
at the highest inclinations. Line profiles in velocity space 
for \civ, \al\ and \mg, are shown in Fig.~\ref{fig:lobal} for a range
of BALQSO viewing angles. We find that ionization stratification
of the wind causes lower ionization material to have a smaller covering factor, 
as demonstrated by figures~\ref{fig:wind} and \ref{fig:lobal}.
This confirms the behaviour expected from a unification model such as Elvis (2000). 
LoBALs are only present at viewing angles close to edge-on ($i>75^\circ$),
as predicted by polarisation results \citep{brotherton1997}.
As observed in a BALQSO sample by \cite{filizak2014}, we find that
BAL troughs are wider and deeper when low ionization absorption features are present,
and high ionization lines have higher blue-edge velocities than the 
low ionization species. 

There is also a correlation between the strength of LoBAL features
and the amount of continuum attenuation at that sightline, particularly
blueward of the Lyman edge as the low ionization base 
intersects the line-of-sight. 
A model such as this therefore predicts that LoBALQSOs and FeLoBALQSOs 
have stronger Lyman edge absorption and 
are more Compton-thick than HiBALQSOs and Type 1 quasars.
An edge-on scenario also offers a potential explanation for the rarity of LoBAL and
FeLoBAL quasars, due to a foreshortened and attenuated continuum, 
although BAL fraction inferences are fraught with complex selection 
effects \citep{goodrich1997,krolikvoit1998}.

\subsection{Parameter Sensitivity}

\label{sec:param_sens}

Having selected an individual fiducial model from the simulation grid, it is important
to briefly explore how specialised this model is, and how small parameter
changes can affect the synthetic spectra. Fig.~\ref{fig:grid}
shows the EW for \civline\ and \mgline\ at a low inclination, 
and $BI$ for \civline\ at a high inclination for the simulation
grid. We find that almost all the models with $f_V=0.1$ are over-ionized, and 
fail to produce strong \civ\ BALs or emission lines. However, 
the models with $f_V=0.01$ generally produce \civ\ BALs and emission lines.
The fiducial model is representative of this family of models and 
the spectra are generally similar, although it is clear from the figure
that we have selected a model near the upper end of the EW distributions.

We find that it is difficult to significantly increase line emission while
keeping the luminosity and mass loss rate of the system fixed.
We show an additional point on figure 7 corresponding to a model with an order of
magnitude higher X-ray luminosity and double the mass loss rate. As expected, 
this results in far higher line EWs, but fails to produce BALs because
the collisionally excited emission swamps the BAL profile. In addition,
this model would lie well above the expected $L_{2kev}-L_{2500}$ 
relation in figure 5. Such a high X-ray luminosity could therefore 
not be the cause of the strong line emission seen in {\em all} Type 1 quasars.

The parameter search presented here is by no means exhaustive, and
we may be limited by the specific parameterisation of the outflow 
kinematics we have used. Nevertheless, we suggest that the angular distribution
of both the line and continuum emission is perhaps the crucial 
aspect to understand. With this in mind, obtaining reliable orientation 
indicators appears to be a crucial observational task if we are to
further our understanding of BAL outflows 
and their connection, or lack thereof, to the 
broad line region. 




\section{Summary And Conclusions}

We have carried out MCRT simulations using a simple
prescription for a biconical disc wind, with
the aim of expanding on the work of H13. To do this, we introduced two main
improvements: First, we included a simple treatment of clumping, and second, 
we improved the modelling of recombination lines by treating H and He as
`macro-atoms'. Having selected a fiducial model from an initial simulation grid,
we assessed the viability of such a model for geometric 
unification of quasars, and found the following main points:
\begin{enumerate}
\item Clumping the wind with a volume filling factor of $0.01$ 
moderates the ionization state
sufficiently to allow for the 
formation of strong UV BALs while agreeing well with the X-ray
properties of luminous AGN and quasars. 
\smallskip
\item A clumpy outflow model naturally 
reproduces the range of ionization states
expected in quasars, due to its stratified density
and temperature structure. 
LoBAL line profiles are seen at a subset of viewing angles, and Fe~\textsc{ii}
absorption is seen at particularly high inclinations. 
\smallskip
\item The synthetic spectra show a \la\ line and weak He~\textsc{ii}~$1640$\AA\ line
as a result of our improved treatment of recombination using macro-atoms. We also see
Balmer emission lines and a Balmer recombination continuum in the optical spectrum, but this
is only really significant at high inclination where the continuum is suppressed.  
\smallskip
\item The higher X-ray luminosity causes a significant 
increase in the strength of the collisionally excited emission
lines produced by the model. 
However, the equivalent-width ratios of the emission lines do not match
observations, suggesting that a greater volume of dense ($n_e\sim10^{10}$~cm$^{-3}$)
material may be required.
\bigskip
\item The line EWs in the synthetic spectra increase with inclination.
BAL and non-BAL quasar composites have comparable EWs, so our model
fails to reproduce this behaviour.
 If the BLR emits fairly isotropically then for a 
foreshortened, limb-darkened accretion disc 
it is not possible to achieve line ratios at low inclinations 
that are comparable to those at high inclinations. 
We suggest that understanding the angular distribution of 
line and continuum emission is a crucial question for theoretical models.
\end{enumerate}
Our work confirms a number of expected outcomes from a geometric unification 
model, and suggests that a simple biconical geometry such as this can come close to 
explaining much of the  phenomenology of quasars. However, our conclusions pose 
some challenges to a picture in which BALQSOs are
explained by an {\em equatorial} wind rising from a classical thin disc, and suggest 
the angular distribution of emission is important to understand if this 
geometry is to be refuted or confirmed. We suggest that obtaining reliable 
observational orientation indicators and 
exploring a wider parameter space of outflow geometries in simulations
are obvious avenues for future work.

\section*{Acknowledgements}

The work of JHM, SWM, NSH and CK is supported by the
Science and Technology Facilities Council (STFC),
via two studentships and a consolidated grant, respectively.
CK also acknowledges a Leverhulme fellowship.
We would like to thank the anonymous referee for a helpful
and constructive report.
We would like to thank Omer Blaes, Ivan Hubeny and Shane Davis for their
assistance with \agn. We are grateful to Mike Brotherton, Mike DiPompeo,
Sebastien Hoenig and Frederic Marin for helpful correspondence regarding
polarisation measurements and orientation indicators.
We would also like to thank Daniel Proga, Daniel Capellupo, Sam Connolly and
Dirk Grupe for useful discussions.  Simulations were conducted using \py\ version 80,
and made use of the IRIDIS High Performance Computing Facility at the
University of Southampton. Figures were produced using the {\tt matplotlib} 
plotting library \citep{matplotlib}. This work made use of the Sloan Digital Sky Survey.
Funding for the Sloan Digital Sky Survey has been provided by
the Alfred P. Sloan Foundation, the U.S. Department of Energy Office of
Science, and the Participating Institutions.

\setlength{\bibsep}{0pt}
\bibliography{paper.bbl}

\end{document}